# Quantifying COVID-19 transmission risks based on human mobility data: A personalized PageRank approach for efficient contact-tracing

**Jiali Zhou**

Department of Urban Planning and Design, The University of Hong Kong, Hong Kong

Address: 8/F, Knowles Building, The University of Hong Kong, Pokfulam Road, Hong Kong

Email: jlzhou@hku.hk

**Zhan Zhao, Corresponding Author**

Department of Urban Planning and Design, The University of Hong Kong, Hong Kong

Musketeers Foundation Institute of Data Science, The University of Hong Kong, Hong Kong

Address: 8/F, Knowles Building, The University of Hong Kong, Pokfulam Road, Hong Kong

Email: zhanzhao@hku.hk

**Jiangping Zhou**

Department of Urban Planning and Design, The University of Hong Kong, Hong Kong

Address: 8/F, Knowles Building, The University of Hong Kong, Pokfulam Road, Hong Kong

Email: zhoujp@hku.hk

# ABSTRACT

Given its wide-ranging and long-lasting impacts, COVID-19, especially its spatial spreading dynamics has received much attention. Knowledge of such dynamics helps public health professionals and city managers devise and deploy efficient contact-tracing and treatment measures. However, most existing studies focus on aggregate mobility flows and have rarely exploited the widely available disaggregate-level human mobility data. In this paper, we propose a Personalized PageRank (PPR) method to estimate COVID-19 transmission risks based on a bipartite network of people and locations. The method incorporates both individuals' mobility patterns and their spatiotemporal interactions. To validate the applicability and relevance of the proposed method, we examine the interplay between the spread of COVID-19 cases and intra-city mobility patterns in a small synthetic network and a real-world mobility network from Hong Kong, China based on transit smart card data. We compare the recall (sensitivity), accuracy, and Spearman's correlation coefficient between the estimated transmission risks and number of actual cases based on various mass tracing/testing strategies, including PPR-based, PageRank (PR)-based, location-based, route-based, and base case (no strategy). The results show that the PPR-based method achieves the highest efficiency, accuracy, and Spearman's correlation coefficient with the actual case number. This demonstrates the value of PPR for transmission risk estimation and the importance of incorporating individual mobility patterns for efficient contact-tracing and testing.

# INTRODUCTION

Since the outbreak of COVID-19, the interaction between the spread of COVID-19 and urban mobility, which describes the movements of people in the urban areas, has received increased attention. On the one hand, the existing scholarship has investigated the impact of pandemics on different travel modes, corresponding mobility patterns, and (changing) attitudes towards these modes. The long-lasting impacts extend beyond service performance and health risks to social equity and sustainable mobility. Barbieri et al. (2021), for instance, after analyzing the impact of COVID-19 pandemic on mobility patterns in 10 countries around the world, showed that there were significant mobility disruptions pertaining all transportation modes and traveling purposes. Dingil and Esztergár-Kiss (2021), using data from an international survey, found a considerable growth in individual transportation modes (e.g., personal vehicles), up to 26% increase for commuting and up to 15% for leisure activities. Public transit ridership, most notably, has decreased dramatically in all metropolitan areas at the early stages of COVID-19 (Dickens, 2020; Jenelius and Cebecauer, 2020). Coupled with the above changes, the number of traffic accidents has decreased, though the average severity of those accidents has seen an increase (Li and Zhao, 2022). On the other hand, there has been emerging scholarship on the role of mobility in the transmission process of infectious disease. Zhu et al. (2021) analyzed the effects of different travel modes and travel destinations on COVID-19 transmission in global cities and concluded that travel modes and destinations can influence the cases in most of the cities in question. Zheng et al. (2020) studied the spatial pattern of COVD-19 transmission via public and private transportation in China and found a significant association between the frequency of flights, trains, and buses from Wuhan, China and daily cases in other cities in China. As a whole, the existing studies have frequently exploited large-scale mobility data to unravel the relationship between the COVID-19 transmission (risk) and human mobility. However, few have quantified transmission risks at the individual level by taking advantage of big data such as transit smart card data, which captures rich and continuous individual-level mobility information for a large sample of the whole population. Quantifying those risks is indispensable for curbing the spread of diseases through efficient (close) contact-tracing and monitoring (Kucharski et al., 2020; Lewis, 2020).

## *COVID-19 and Contact-tracing*

In the field of epidemiology, contact-tracing is used as a target control strategy to identify people who have been in close contact with an infectious individual (Klinkenberg et al., 2006). In reality, it is often challenging to capture all close contacts due to limited human resources (Lewis, 2020). Manual contact-tracing is slow and labor-intensive and it relies heavily on the infected person's ability to recall his/her past experiences. Digital contact-tracing (DCT) was introduced to overcome the aforementioned shortcomings and limitations of manual contact-tracing (Kucharski et al., 2020a). Most of them use Wi-Fi or Bluetooth technology (Yang et al., 2021) and transit smart card (Park et al., 2020). DCT is effective in identifying close contacts, but as the number of infections increases, the number of close contacts can become very large, one infector can have as many as 200 close contacts in reality (Lewis, 2020). According to the WHO's benchmark, a COVID-19 contact-tracing operation needs to trace and quarantine 80% of close contacts within 3 days of a case being confirmed to contain the spread (World Health Organization, 2020).

The limitations of current contact-tracing methods require more precise and efficient strategies to allocate and relocate currently available resources so that local governments can improve the speed and quality of contact-tracing and slow the spread of infectious diseases such as COVID-19. Despite the vast array of studies in contact-tracing and mass testing strategies (Kucharski et al., 2020b; Quilty et al., 2021), with some of which based on symptoms of people, few have attempted to compare the efficiency of different methods of transmission risk estimation and contact-tracing strategies. Transmission risk analysis helps quantify risk levels among people and locations in cities, laying the groundwork for prioritizing resources to the most risky people and hotspot areas in a more targeted way.

*COVID-19 Risk Analysis*

To track the contacts and contain the spread of COVID-19 more efficiently, areas and sub-populations with higher transmission risks need to be identified so that the available resources can be prioritized accordingly. For example, governments around the world developed standards for characterizing areas such as streets or counties based on low, medium, and high risks, and designed different testing, quarantining, and isolation strategies for areas of different risk levels (Centers for Disease Control and Prevention, 2022; Chinese Center for Disease Control and Prevention, 2020). Researchers have proposed models at different scales and resolutions to measure risks, identify hotspots, and model transmission processes. For example, Bird et al. (2020) measured and classified different countries' risks into four risk categories based on COVID-19 data. However, these studies focus on the population in larger areas, such as country, city, or county level, and consider the transmission risk over a time horizon of days, weeks, or months. To achieve the prevention and containment of infectious diseases in a more precise way, a fine-grained person- or location-based risk estimation framework is required.

The contact network model is widely applied in infectious disease studies for more detailed transmission modeling and contact-tracing. For example, Liu et al. (2020) employed a more realistic contact network model to reconstruct contact and the airborne spread to model the outbreak of COVID-19 on the Diamond Princess cruise ship. They argued that the identified high-risk susceptible persons should follow stricter control and prevention measures. Mo et al. (2021) proposed a time-varying weighted encounter network to model epidemic spreading within public transportation systems and concluded that isolating influential passengers at an early stage can reduce the spreading. These models concentrate on the interactions between people in a specific area or transportation mode, though most infections are likely to occur at a large scale across different neighborhoods and communities. On the other hand, some network models for locations often ignore the interactions among individuals and interactions between individuals and locations. For instance, Bhattacharya et al. (2021) constructed a network of various provinces and cities and applied the PageRank algorithm to identify hotspot provinces and cities in India. Chang et al. (2021) introduced a metapopulation susceptible–exposed–infectious–removed (SEIR) model that integrates fine-grained, dynamic mobility networks to simulate the spread of SARS-CoV-2 in ten of the largest US metropolitan areas. They found that a small minority of 'super-spreader' points of interest account for a large majority of the infections. Zhou et al. (2022) proposed a station risk classification method considering the network structure and aggregate passenger flow, but ignoring the interactions at the individual level. Yabe et al. (2022) utilized a contact network based on human mobility trajectories (GPS traces) and web search queries to predict COVID-19 hotspot locations. They also proposed a high-risk social contact index to capture contact density and COVID-19 contraction risk levels. These studies, often employ unipartite networks of people or locations, ignoring the interactions between them.

The transmission processes take place through contact networks of infected individuals. However, due to the uncaptured interactions between people in the contact networks and at different locations, the specific transmission process is highly uncertain and can only be estimated to some extent. Generally, people who visited high-risk locations have a higher possibility to be infected, and locations that are visited by many high-risk people tend to be high-risk as well. It is important to consider both individual mobility traces across locations and colocation patterns across individuals. In this paper, we show that this can be done by constructing and analyzing a heterogeneous bipartite network that connects people and their visited locations using large-scale human mobility data.

The overarching objective of this paper is to develop a network-based approach to quantifying transmission risks of infectious diseases (e.g., COVID-19) across people and locations. It leverages the widely available disaggregate-level human mobility data, adapts the well-known PageRank (PR) algorithm for analysis of the individual mobility networks, and can be used for efficient contact-tracing given the source of infection. Both synthetic data and real-world transit smart card and COVID-19 data from Hong Kong are used to validate the proposed methodology. The specific contributions of this paper include:

- Based on the mechanism underlying the spread of infectious diseases (e.g., COVID-19), we demonstrate how disaggregate-level human mobility data can be represented as a bipartite network between people and their visited locations for estimation of transmission risks.
- To incorporate individual mobility patterns, we propose a Personalized PageRank (PPR) approach to estimating the transmission risks across people and locations in the aforementioned bipartite people-location network. The proposed PPR approach captures the source of infection, which could be a specific person or location.

- Based on extensive experiments using synthetic and real-world data from Hong Kong, we show that the proposed method can lead to a more accurate estimation of COVID-19 transmission risks and more efficient contact-tracing and testing, compared to other commonly used strategies.

## METHODOLOGY

### *Research Framework*

The overall methodological framework of the study is described in Figure 1 below. Disaggregate-level human mobility data, such as transit smart card data, is used for bipartite people-location network construction, based on which PPR scores are calculated for transmission risk estimation. Based on the PPR scores, several contact-tracing and testing strategies are designed. We explored the validity and effectiveness of the proposed method using both microscopic transmission simulation for individual-level evaluation and real-world pandemic data for neighborhood-level evaluation. Different mass tracing/testing strategies are compared, highlighting the efficiency of our proposed method.

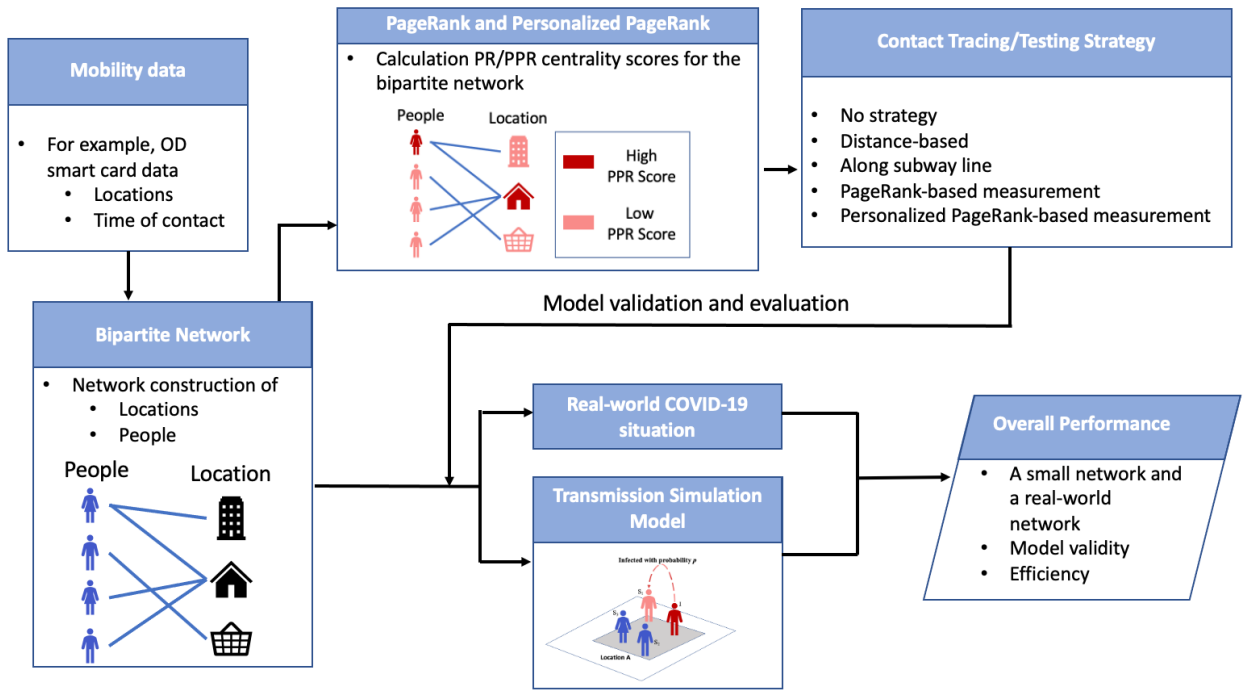

**Figure 1. Overall methodological framework**

*People-location Network Graph*

As mentioned before, unipartite models are not enough to capture the typical relationship between locations and people in urban mobility systems. In real-world applications, not only should the relationships between entities of the same type be considered, but the relationships between different types of entities should also be modeled.

While most existing network models are built on a homogeneous network of locations or people, we propose a heterogeneous people-location network that connects people and their visited locations in a bipartite graph. This is particularly useful in disease transmission risk analysis, as people can carry a virus across locations and the virus can survive at a location (even without a living host) for a certain time. For example, the SARS-COV-2 virus (the virus that causes COVID-19) can survive on a surface for days (Van Doremalen et al., 2020).

The bipartite graph is a ubiquitous data structure that can model the relationship between two entity types: for instance, people and locations. In a bipartite graph, all vertices are characterized into two sets, within which vertices in the same set cannot be directly connected, but can only be connected via a vertex from another set. Based on increasingly prevalent human mobility data, such as cell phone data and smart card data, the movement of people can be described as a bipartite network with nodes representing people and locations, and each link signifies an individual visiting the location. The detailed illustration of the construction of a people-location network can be found in Supplemental Materials Table S1 and Figure S1.

In this study, we construct a people-location network based on human mobility in a limited time period. While the temporal dimension is important for risk analysis of epidemics, the time-evolving nature of epidemics varies, depending on different cities' urban forms, socioeconomic characteristics, and local policies. These contexts impact the scalability of a risk estimation method. Instead of capturing the time-evolving nature, this study focuses on the initial stages of an outbreak, providing a tool for policymakers to make predictive rather than reactive management decisions. Our proposed methods use only a short time period's (one day's) human mobility data to help with more proactive mass tracing/testing. The reason is that people will be contagious only around 2.7 days after being exposed to the original SARS-COV-2 virus (Schive, 2020), and this period differs among later variants. It is difficult to model and capture the $2^{nd}$, $3^{rd}$ generation infection cases accurately. The number of potential interactions among people increase exponentially as the time window expands. Therefore, our study focuses only on the starting phase of a pandemic and the $1^{st}$ generation of infection cases for fast transmission risk early-warning.

*PageRank and Personalized PageRank Algorithms*

Brin and Page (1998) proposed the original PageRank algorithm for measuring webpage importance. Suppose we have webpage A linking to another webpage B, considered as A 'voting' for B.

The importance of webpages is based on the number of votes a webpage can obtain. The voting is also weighted by the importance of related webpages. For example, if webpage A is important (with high PageRank values), webpage B will also have an increased PageRank value because it is pointed to by A.

The algorithm begins by assigning an equal initial PR score to all nodes. Then, in each iteration, the PR score of each node is recalculated based on the number and quality of their links in the network. The PR score converges after a sufficient number of iterations. The calculation is conducted by dividing the PR score of the source node by the number of its out-links. The PageRank score of a node $p$ is given as:

$$PR(p) = \frac{1-d}{N} + d \sum_{i=1}^{k} \frac{PR(p_i)}{C(p_i)} \tag{1}$$

where $N$ is the total number of nodes in the network, $p_i$ is the node that links to node $p$, $d$ is a damping factor, and $C(p_i)$ is the number of out-links of $p_i$. The damping factor $d$ defines the probability of a web surfer going to an adjacent link, and thus the probability of a surfer skipping to another page is then given by (1- $d$).

The original PR algorithm has been introduced to identify the high-risk locations for COVID-19 transmission (Bhattacharya et al., 2021), but it is only based on aggregate mobility flows and a homogeneous network of locations. In this study, we propose to apply an adapted version of the PR algorithm, called Personalized PageRank (PPR) to capture the interactions between people and their visited locations, and account for the effects of the originating individuals or locations.

The PPR algorithm was proposed by Page et al. (1999) to model the reachability of a node $t$ in the network with respect to a source node $s$. Suppose we have a source node $s$ and a target node $t$ in the network, the PPR value $\pi(s, t)$ is then the probability that a random user from $s$ will reside at $t$, which represents the importance between $s$ and $t$.

The PPR score calculation can be expressed as:

| | Algorithm 1 PPR Algorithm |
|---|---|
| 1 | Input Graph $G$ with $N$ nodes |
| 2 | Initialize damping factor $d$ |
| 3 | Initialize personalized node seeds, specifying the source node $s$ in $G$ |
| 4 | Initialize all nodes in $G$ with original PR score = 1/$N$ |
| 5 | While not converged |
| 6 |    For all node $p$ in the graph, do |
| 7 |       **If $p$ is source node $s$** |
| 8 |          $PPR(p) = 1 - d + d \sum_{i=1}^{k} \frac{PR(p_i)}{C(p_i)}$ |
| 9 |       Else |

| | | |
|---|---|---|
| 10 | | $PPR(p) = d \sum_{i=1}^{k} \frac{PR(p_i)}{C(p_i)}$ |
| 11 | End for | |
| 12 | If error rate for any vertex in the graph falls below a given threshold, | |
| 13 | Converged | |
| 14 | End While | |

PPR has been used for personalized web search and recommendation (Gleich and Polito, 2007). PPR was also utilized in the social network studies to analyze the network structure and the importance of each user (Garcia et al., 2013; Pedroche et al., 2013). Personalized PageRank can be applied to a bipartite network to decide the importance of vertices in two different sets of nodes (Kłopotek et al., 2016). However, few studies utilized PPR in the urban networks or transportation networks to study the importance of nodes (location nodes and person nodes).

In this study, we adapt PPR for a bipartite human mobility network to quantify COVID-19 transmission risks across different locations and people, so that we can identify high-risk areas and individuals for efficient contact-tracing and mass testing. Unlike PR, PPR is better suited for our problem because it allows us to capture the source of the infection. The adapted PPR score measures the importance of location and people nodes relative to the source of infection, which can be the basis for more efficient resource allocation and prioritization.

*Simulation Framework*

It is difficult to evaluate our proposed approach using real-world data due to the lack of detailed personal and spatiotemporal epidemiological information about COVID-19 transmission traces. As a result, this study adopts the microscopic agent-based simulation method of the SEIR model for transmission process modeling, which was proposed by Okabe and Shudo (2021) to model the transmission process, and use the simulation result as the ground truth to evaluate the performance of our proposed method. The process is below:

1. A network of people and locations is generated
2. At the beginning of the simulation run (*t=0*), an 'originating' individual (I) is infected.
3. The *k* susceptible individuals (S) around the 'originating' individual (I) will be infected (I) with a probability *p*.
4. Repeat step 3, until the 'originating' individual (I) is isolated and cannot enter the mobility system any more.
5. The time sequence obtained from the above procedures is seen as a replication. The simulation will run for a specified number of replications.

The transmission process simulation can be described using the example bipartite network of people and location shown in Figure 2. The original infected person is *I*, and it shares space with susceptible persons $S_1$ and $S_2$ at location 1 at Time $t_0$. $S_1$ and $S_2$ are close contacts of *I*, and with probability *p*, $S_1$ will be infected by *I*. As a result, $S_1$ becomes $I_1$ at the next time step $t_1$. The objective of simulation is to compare the cost effectiveness of different strategies in finding the close contacts of the source person of infection (i.e., patient zero) for fast testing and treatment, and quickly identifying secondary COVID-19 infection cases from the source person of infection.

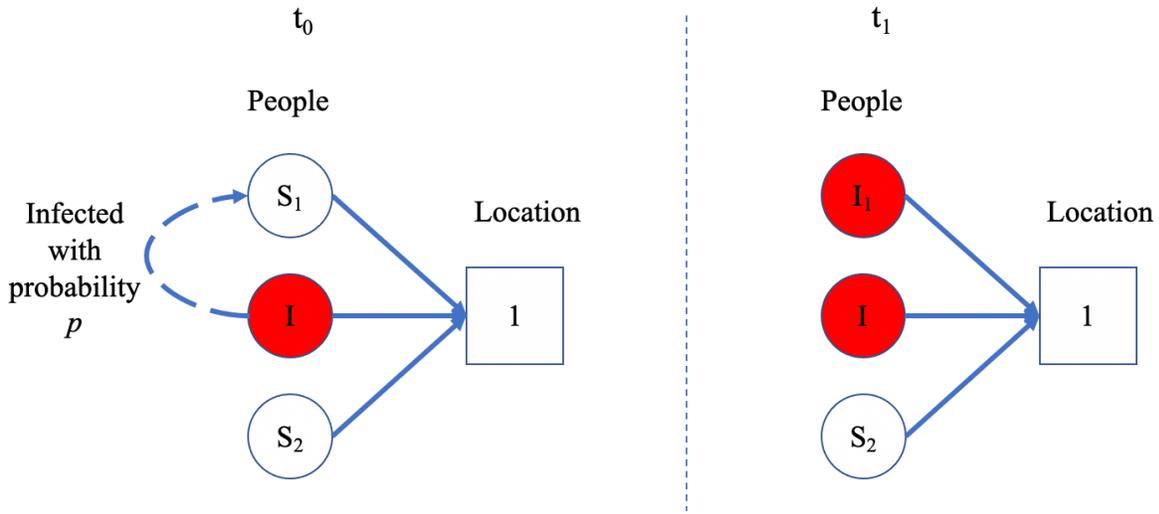

**Figure 2. Simple illustration of transmission process**

*Contact-tracing and Mass Testing Strategies*

According to WHO Headquarters (HQ) (2021), the mass COVID-19 testing should address the use of available resources to maintain the highest level of public health impact. In this paper, we compare various strategies to highlight the performance of our proposed PPR-based method in quantifying transmission risks for people and locations and prioritizing available resources for tracing/testing close contacts of the infected individuals and locations. The strategies are illustrated using the example urban system in Figure 3 below, where location A is the originating location of an ongoing outbreak. The government can take different measures to trace and test all possible close contacts based on their estimated risks. The potential strategies include:

1. **Base case (No strategy)**. Under the no strategy case, the government tests and tracks as many people as it can to control the spread, but without differentiating the people and locations in the cities regarding transmission risk. For example, if the government's capacity can cover 10% of the population, then it will arbitrarily select 10% of the population in the urban area for contact-tracing and testing. As shown in Figure 3a, users A, B, and C will be treated without any differences with regard to being tested or traced.

2. **Location-based strategy**. Given the source individual or location of the outbreak, the location-based strategy prioritizes resources to the people closest to this originating location, as advised by WHO Headquarters (HQ) (2021). As shown in Figure 3b, user A has a higher priority to be traced and tested since it is closer to the originating location of the outbreak.

3. **Route-based strategy**. The route-based strategy assumes that virus dissemination tends to take place along a specific transportation route. Some local governments in China used this method for containment of the COVID-19 spread such as the city of Dalian, China (Agence France-Presse, 2020). Under this strategy, the locations along and close to a specific transportation route are prioritized. As shown in Figure 3c, there are two major transit routes 1 and 2, user A has a higher priority to be traced and tested since it is closer to route 1, which passes right through location A.

4. **PR-based strategy**. Under this strategy, a bipartite graph of visited locations and people will be constructed. A PR score will be calculated for every node in the network, including location nodes and user nodes. The available resources will be allocated according to the PR scores of the people. As shown in Figure 3d, the color-coded user A and B will have more priority to be traced and tested since they have higher PR scores.

5. **PPR-based strategy**. Given the originating location of the virus spread or the specific originating person, we set the location or the person as the source node in the algorithm. A PPR score will be assigned to every node in the network, including all locations and people. The available resources will be allocated according to the PPR scores of the people. As shown in Figure 3e, assuming location B is the source of the outbreak, users C and B have more priority to be traced and tested since they have higher PPR scores.

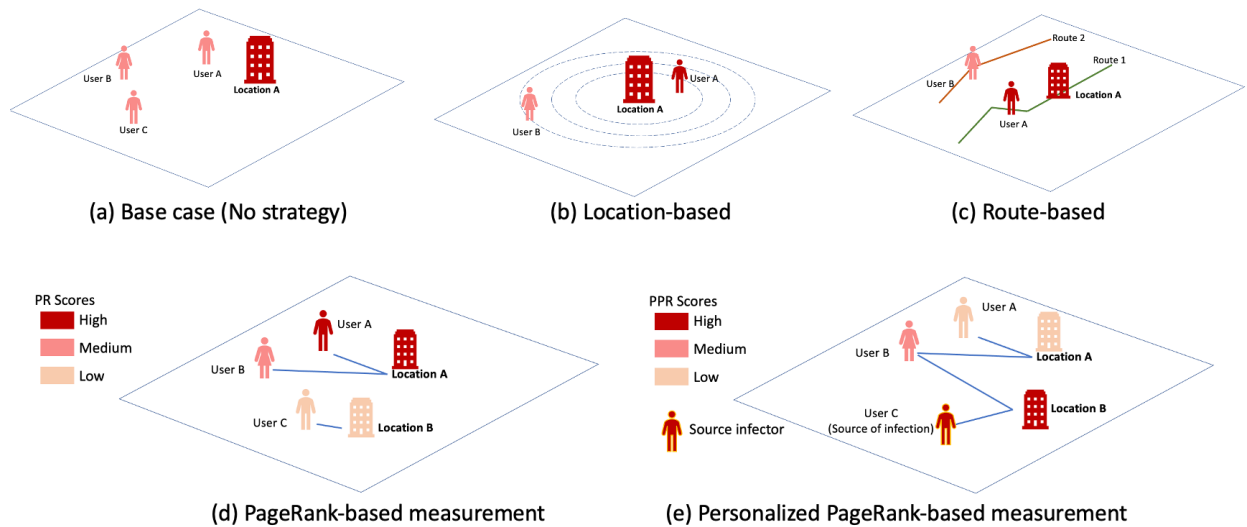

**Figure 3. Contact tracing and testing strategies**

*Experimental Design*

To evaluate the proposed approach, the following networks were tested:

1. **A synthetic network of human mobility**, including the travel history of 20 people and 3 locations they visited.
2. **A real-world network of human mobility**, based on transit smart card data provided by the Hong Kong Mass Transit Railway (MTR).

MTR is a major public transport network serving Hong Kong with the highest frequency and daily ridership (Legislative Council Secretariat of Hong Kong SAR, 2016). The system included 230.9 km of rail with 98 stations. It is worth noting the MTR system accounts for a large portion of all trips in Hong Kong. Specifically, the smart card data from January 21, 2021, consisting of 6.3 million transaction records, is used to represent the human mobility pattern of a typical day. Figure 4 shows human mobility pattern as revealed in the data. Each connection represents one trip, while the red end represents the origins and the blue end destinations.

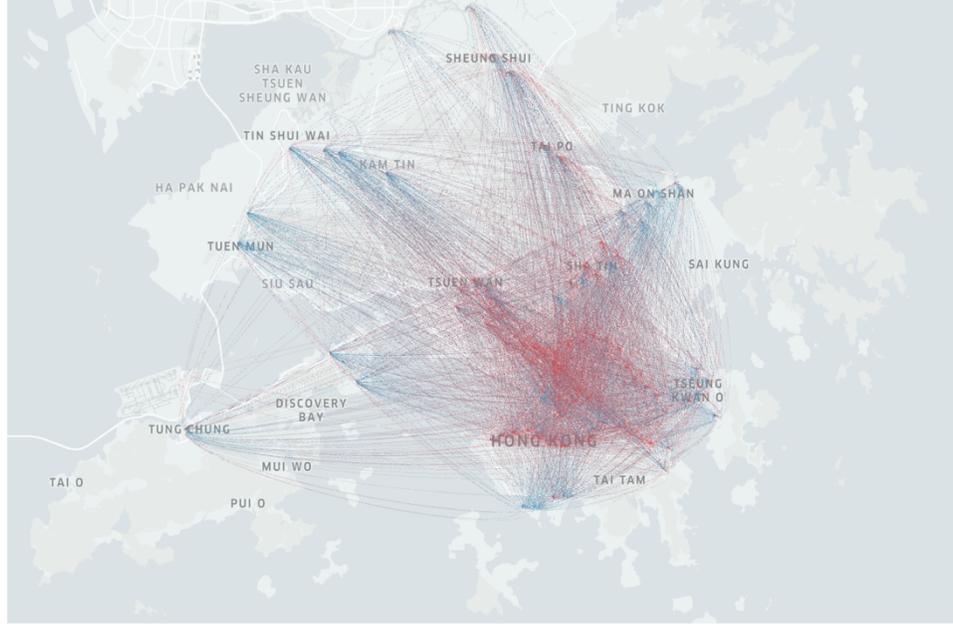

**Figure 4. Human mobility pattern based on MTR data**

For individual-level evaluation, due to the lack of detailed investigation on the infection, the simulation is used as a testbed for contact-tracing and importance ranking strategies in identifying secondary COVID-19 infection cases from the source person of infection (patient zero). $\beta$ is defined as the rate of infection, and we have:

$$\beta = kp \quad (2)$$

where $p$ is infection probability and $k$ the number of close contacts. According to Okabe and Shudo (2021), $\beta$ is set as 0.4 for COVID-19's rate of infection.

For neighborhood-level evaluation, the entire city is divided into 291 Tertiary Planning Units (TPUs), and the number of actual COVID-19 cases is summarized for each TPU as the actual transmission risk. If the TPU contains MTR stations, the PR and PPR scores for MTR stations in the TPU are aggregated for that TPU area as the predicted transmission risk.

*Performance evaluation metrics*

The performance of various mass tracing/testing strategies is assessed using a number of metrics from the individual and neighborhood-level points of view:

1. **Recall** is used as the main individual-level evaluation metric to assess the efficiency and validity of the strategies. In our case study, we define the recall as:

$$Recall = \frac{N_c}{N_p} \quad (3)$$

where $N_p$ is the number of people being infected (positive) by the source of the infection in the simulation and $N_c$ is the number of people being infected (positive) and, at the same time, being identified as the prioritized group for mass tracing and testing by the proposed strategies. A perfect strategy has a recall equal to 1. However, recall by itself does not consider efficiency (or precision), therefore, we will calculate recall under different testing capacity assumptions. A good contact-tracing/testing strategy should be able to achieve good recall even with limited testing capacity.

2. **Accuracy** is a neighborhood-level evaluation metric, defined as the percentage of the confirmed cases (in reality) in the areas which are predicted to be high transmission risk areas. The high transmission risk areas are the areas with a PR or PPR score higher than the 80[th] percentile value, as introduced by Tong et al. (2022).

3. **Spearman's correlation coefficient** measures the strength and direction of association between two ranked variables $X$ and $Y$, is used for model validation.

$$r = 1 - \frac{6 \sum d^2}{n(n^2-1)} \tag{4}$$

where $d$ is the difference between the two ranks of each observation and $n$ is the number of observations. A value of 1 implies that $Y$ always increases as $X$ increases. A value of −1 implies that $Y$ always decreases as $X$ increases. A value of 0 implies that there is correlation. Spearman's correlation coefficients between the ranked number of cases and ranked estimated risk scores are compared among the aforementioned strategies.

**RESULTS**

In this section, various experiments are implemented using synthetic and real-world urban mobility datasets. We investigate the efficiency and effectiveness of different mass tracing/testing strategies, including strategies based on the proposed PPR approach, in capturing potential secondary COVID-19 infection cases.

*Model evaluation based on synthetic mobility data*

In this section, we test the effectiveness of the proposed mass tracing and testing strategies on capturing the potential infectors, from a microscopic and individual perspective. To do this, we generate a small synthetic mobility dataset of 20 individuals across 20 locations. The specific visiting history can be found in Supplemental Materials Table S2. The synthetic data is used to demonstrate the construction of the people-location networks, and application of the PR and PPR algorithms, and how we evaluate the efficiency of different strategies.

Based on the synthetic data, we construct a people-location bipartite network that connects the 20 individuals (No. 1-20) and 3 locations (A, B, C). The people node's average degree is 2, and the location node's is 13.3. Each link between an individual and a location indicates that the individual has visited the location. As an example, we assume the originating infector to be person No.18, and he has visited location C only. The red color represents the original infector and the location he visits. Person No.18 is treated as the source node in the PPR score calculation. According to the common measure of some cities, other people who visited location C at the same time as person No.18 and location C are deemed as high-risk (Los Angeles County Department of Public Health, 2022; the Government of the Hong Kong Special Administrative Region, 2021).

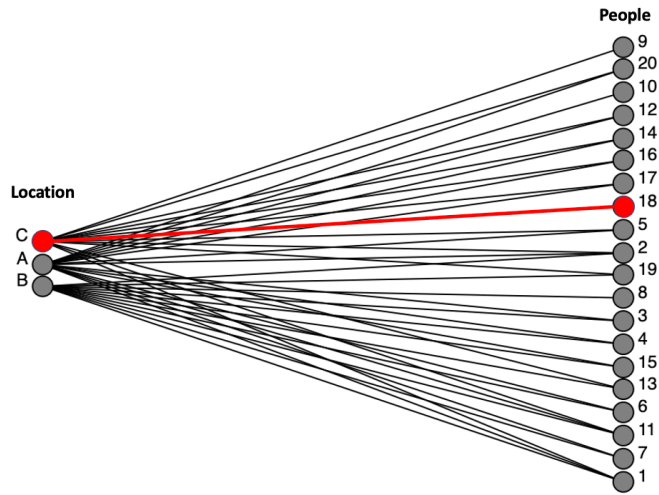

**Figure 5. A Small Synthetic People-Location Network**

*Microscopic simulation of transmission among people*

A microscopic simulation is used for the disease transmission process due to a lack of detailed epidemiological data. We simulate the virus transmission and spread among individuals in the people-location network for 1000 replications. In each replication, person No.18 is selected as the source of infection, but a different set of individuals would end up being infected. Through multiple replications, the number of times each user gets infected is calculated. For model validation, we will treat the simulated transmission results as the "ground-truth" and see whether the PR and PPR can effectively identify the infected individuals (from the simulation) as high risk.

*Individual-level analysis*

PR and PPR scores for locations and people are calculated for risk estimation in the people-location network. The PR and PPR scores are shown in Figures 6a and b. The relative PR and PPR scales are shown for comparison. The PR concerns more about the number and quality of links. In the small network example, PR can identify the most visited location (location A), and most active users. However, location C visited by the infector is not identified as the highest risk location by PR. Also, many users who visited location C were not categorized as high-risk contacts. The PPR method treats person No.18 as the source node, and other locations and people as the target nodes. It can identify the source of infection and the transmission via the bipartite mobility network. Some location C's visitors visited location A, which brings risk to location A. In general, from a qualitative perspective, PPR can capture the high-risk areas and people more accurately than PR, which identifies only the nodes with more links and more high-weight links.

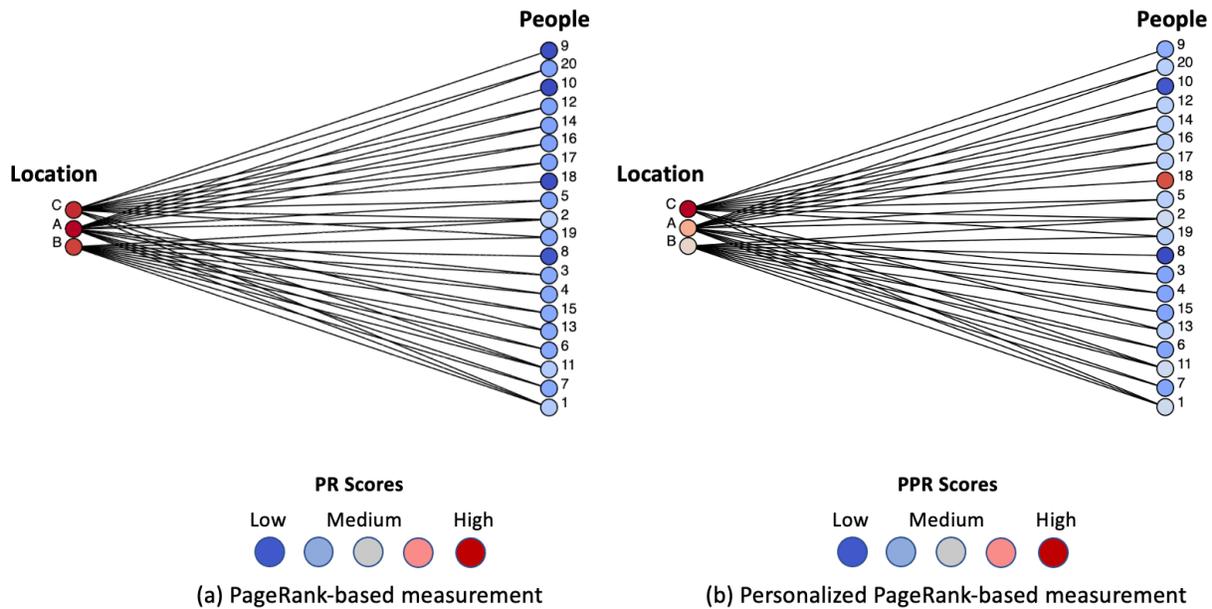

**Figure 5. PR and PPR scores for transmission risk**

To further examine PR and PPR's ability to capture potential high-risk locations and people quantitatively, we compare the recall for evaluating the efficiency from an individual perspective of different mass tracing and testing strategies. The strategies include (1) base case (no strategy), (2) location-based, (3) route-based, (4) PR-based, (5) PPR-based. In the synthetic network, the geographic factors are not considered, so the location-based and route-based strategies are not considered for the comparison.

The recalls from strategies (1), (4), and (5) are compared under different assumptive levels of testing capacity of the local municipality. The contact-tracing/testing capacity of the local municipality is expressed as the percentage of the people in the related population that can be tested with the available resources.

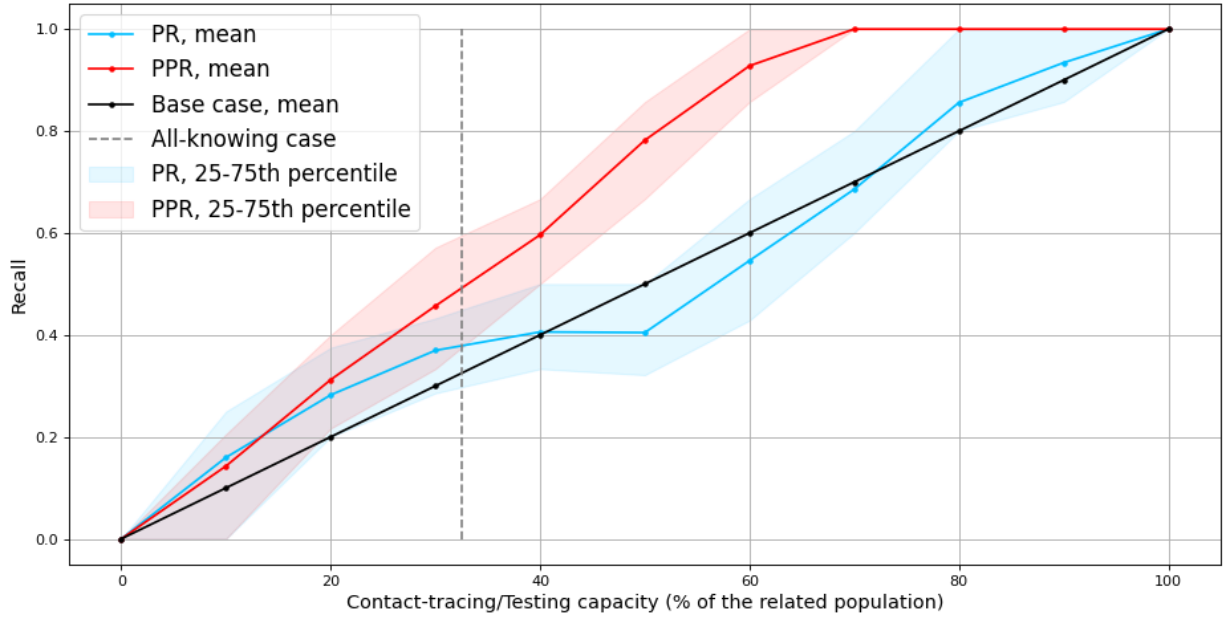

**Figure 6. Recall for PR- and PPR-based strategies under various testing capacity levels**

Figure 6 shows the recall values under different levels of tracing and testing capacity. The all-knowing case is the situation where we know exactly who the secondary infectors are, which provides an upper bound to different strategies. The x-axis represents different levels of tracing and testing capacity, ranging from 10% to 100%. The required tracing/testing capacity for the all-knowing case is around 33%. It is clear that the recall value of the PPR-based strategy is significantly higher than the PR-based strategy and the base-case strategy. The efficiency of the PPR-based method is also closer to the all-knowing case. The PR-based strategy is having a lower value of recall compared with PPR-based strategy, and in some cases, the PR-based method has even lower recall than the base case, partly due to its focus on links and link quality, instead of the source of the infections. The popularity of the locations does not necessarily decide the risk levels.

*Model evaluation based on Hong Kong MTR data*

In this section, we present a case study based on real-world large-scale mobility data from Hong Kong MTR, with the aim to thoroughly compare different types of strategies. Based on the data, we construct a bipartite network of 98 location nodes and 1.7 million individual nodes. The individual node's average degree is 3.6, and location node's is 86163.3. With the location-people network built, we can assume either a person or a specific location as the source of the outbreak in the PPR algorithm. At the time of writing (February 2022), Hong Kong is in the middle of the 5$^{th}$ wave of COVID-19 outbreak, and the neighborhood of Tuen Mun (circled area in figure 8a) is a major source of early cases starting from late

December 2021. (Hong Kong S.A.R. Government, 2022a). Thus, to better approximate the real-world pandemic situation, we choose Tuen Mun MTR station as the source node in our PPR analysis.

*Microscopic simulation of transmission among people*

Due to the lack of detailed personal epidemiological investigation data about infectors' trajectories, we simulate the virus transmission and spread among users in the people-location networks for 1000 replications, and the number of times each user is infected is calculated. In each replication, a visitor to Tuen Mun station is selected randomly as the source of infection in the simulation, and the transmission will start from this individual.

*Individual-level analysis*

Similar to the synthetic data case, due to the lack of detailed personal epidemiological investigation data about infectors' trajectories, we simulate the virus transmission and spread among users in the people-location networks for 1000 replications, and the number of times each user is infected is calculated. In each replication, a visitor to Tuen Mun station is selected randomly as the source of infection in the simulation, and the transmission will start from this individual. For individual-level evaluation, we focus on whether the actually infected individuals will be identified as high-risk by the proposed algorithms and strategies.

Again, we compare the recalls under different testing capacities, for evaluating the efficiency of different mass tracing/testing strategies. The strategies analyzed included (1) base case (no strategy), (2) location-based, (3) route-based, (4) PR-based, (5) PPR-based. The results are shown in Figure 7. It is apparent that the PPR-based strategy has the highest efficiency in mass tracing and testing, as compared to other strategies. 5% of local government tracing and testing capacity is sufficient to capture all secondary infections. The efficiency of the PPR-based method is also closer to the all-knowing case.

Different from the small synthetic network result, the PR-based strategy is more efficient than the base case under all levels of contact-tracing/testing strategy. The recall of the PR-based strategy is higher than the route-based strategy when the tracing and testing capacity is between 50% and 80%. This is because the PR-based strategy captures the active users and those who go to the popular locations, which also corresponds to the high-risk people from the simulation. Another reason is that the network structure and average node degrees differ between real-world and synthetic networks. The location-based and route-based strategies are less efficient than PPR-based strategy. The location-based method captures local interactions in nearby neighborhoods and the source of the outbreak to some extent. On contrary, the route-based strategy fails to capture the passenger interactions from other lines, while transferring is very common in the Hong Kong MTR system.

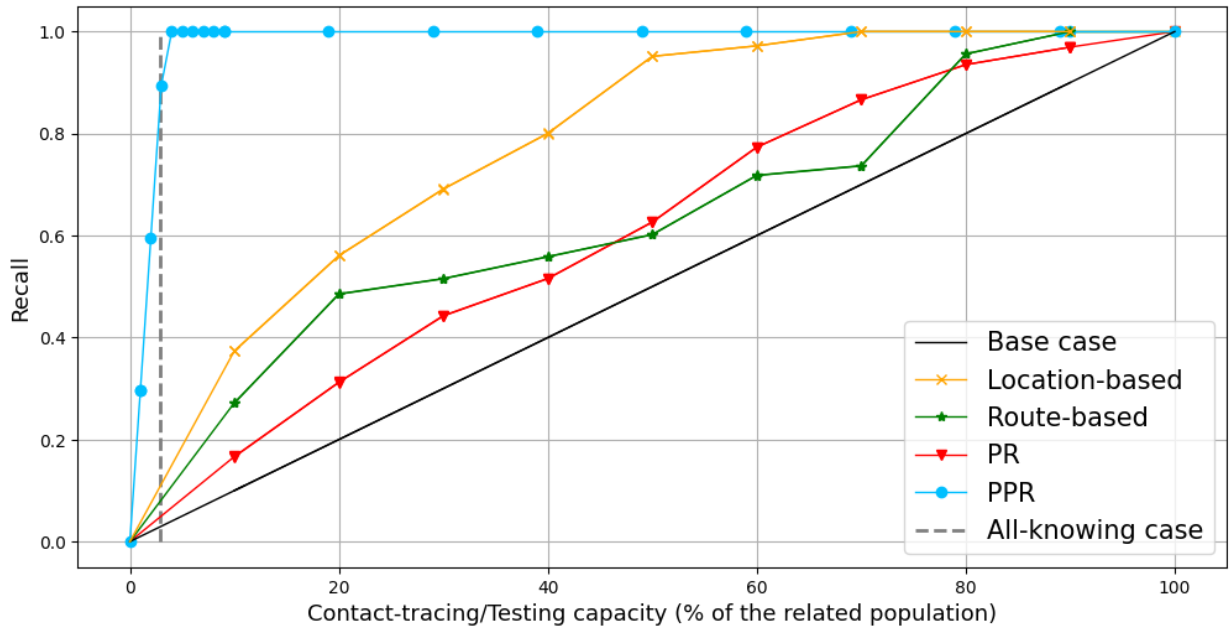

**Figure 7. Recall of base case (no strategy), location-based, route-based, PR-based, PPR-based mass tracing/testing strategies**

*Neighborhood-level analysis*

        As it is difficult to validate the model using personal epidemiological reports from reality, accuracy and spearman's correlation are employed to evaluate model performance at the neighborhood-level. We collect real-world pandemic data, including the number and locations of COVID-19 cases, and compare with the proposed risk estimation method. The fifth wave of COVID-19 outbreak starting from December 2021 in Hong Kong (Hong Kong S.A.R. Government, 2022b) provides the "ground truth" for us to validate the strategies and algorithms using real data. In Figure 7, we compare the actual high-risk areas, the PPR- and PR-predicted high-risk areas for all TPUs.

        The results in Figure 8 shows that the PPR-predicted high-risk areas (Figure 8c) correspond more with the reality (Figure 8a), though PPR misses some high-risk TPUs around the international airport. In normal operations, people traveling to those TPUs are not served by the public transportation system and thus not captured in the MTR data. Therefore, the transit smart card data cannot capture the risk at those TPUs. On the other hand, the PR-predicted areas are mostly located in the densely populated neighborhoods, while missing the source of the fifth wave outbreak (the circle area in Figure 9a).

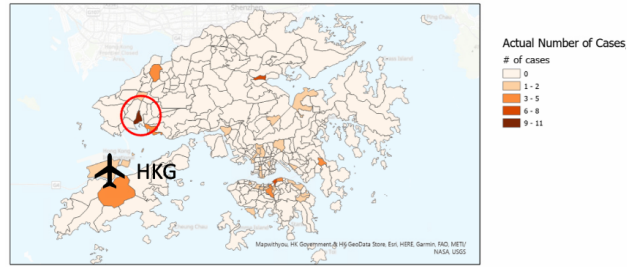

(a) Reality High Risk areas (up to Jan 11, 2022)

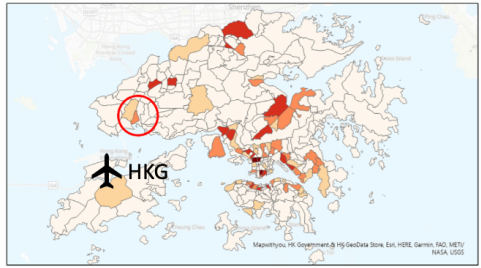

(b) PR High Risk areas

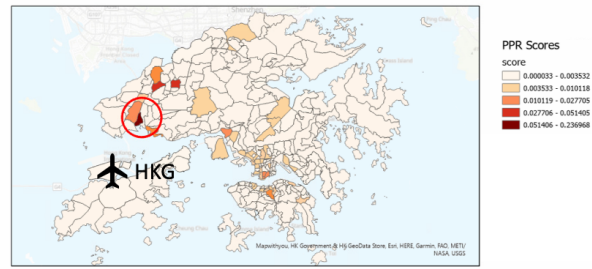

(c) PPR High Risk areas

**Figure 8. High-risk areas in reality, in PR and PPR methods**

Table 1 compares the accuracy and Spearman's correlation coefficients under different strategies. The location-based strategy assesses the risk based on distance from the source of the outbreak, while the route-based strategy assesses the risk based on the route of the MTR line. Accuracy is used to evaluate the algorithms and strategies' efficiency from the neighborhood-level perspective. Be reminded that accuracy in our study is defined as the percentage of the confirmed cases (in reality) in the areas which are predicted to be high transmission risk areas. The PR- and PPR-based strategies have similar accuracy, and location-based (52%) is better than the route-based (32%) strategy. The reason for the similarity between PR and PPR-based strategies in the neighborhood-level analysis could be PR-based strategy identifies the areas with high passenger volume, which are often the densely populated areas with higher transmission risks in the reality. The results show that the PR and especially the PPR algorithms can serve as tools for transmission risk prediction in cities, capturing the high-risk areas with an accuracy of 60%. Spearman's correlation coefficient is used to show the correlation between predicted transmission risk ranking and the actual number of infection cases for different TPUs. The results show that the PPR-based method has the highest correlation with reality, confirming the superiority of the proposed method.

**Table 1. Accuracy and Spearman's correlation coefficient**

|  | Accuracy | Spearman's correlation coefficient |
|---:|:---:|:---:|
| Base case | 20% | 0.01 |
| Location-based | 52% | 0.06 |
| Route-based | 32% | 0.01 |
| PR-based | 60% | 0.31 |
| PPR-based | 60% | 0.33 |

**CONCLUSION AND DISCUSSION**

In this paper, we discuss the problem of how to estimate the transmission risks of infectious diseases (e.g., COVID-19) across people and locations, and how to deploy efficient contact-tracing and treatment measures, using the disaggregate-level human mobility data. Specifically, we first represent human mobility as a bipartite network of locations and people, and then adapt the Personalized PageRank (PPR) algorithm for COVID-19 transmission risk estimation, incorporating individual mobility patterns and their interactions. The empirical validation of our approach is based on both a synthetic mobility network and a real-world mobility network using smart card data from Hong Kong MTR. According to the comparative results at the individual and neighborhood levels, the proposed PPR-based approach can estimate the transmission risks of people and locations effectively and accurately, and used for mass tracing and testing under capacity constraints. PR-based, location-based, and route-based strategies are less effective in identifying high-risk individuals or locations. The PR and PPR metrics and the associated tracing and testing strategies can be used for an early warning system to identify potential high-risk people and locations, as well as personalized alerts on mobile apps for high-risk people. Furthermore, the identified high-risk areas can be the target for targeted sewage surveillance for rapid containment of any outbreak.

The main contributions of this study are (1) the representation of human mobility as a people-location bipartite network between people and their visited locations for estimation of transmission risks, (2) a PPR approach, which captures the source of infection, to estimate the transmission risks across people and locations in the aforementioned bipartite network, and (3) empirical validation of the proposed approach, showing that the proposed PPR-based method can lead to more accurate estimation of COVID-19 transmission risks and more efficient contact-tracing/testing strategies, compared with other more commonly used strategies.

Some limitations of the study include the negligence of time as an important factor for transmission modeling. Future studies can focus on developing PPR algorithms with a time-inheriting characteristic, which captures the development of the outbreak. The PPR algorithm for transmission risk measurement could also consider different locations' socioeconomic characteristics, and the points of interest around that area to consider the local characteristics and contexts.

**Supplemental Material**

To illustrate the construction of a bipartite network of people and locations, let us consider a simple urban mobility network consisting of 10 people (1, 2, 3, …, 10) and 4 locations (A, B, C, and D).

As shown in Table S1, the 10 people visited the four locations at different time, $t_1$, $t_2$, $t_3$, and $t_4$ (and $t_1 < t_2 < t_3 < t_4$).

Table S1. Travel history of 10 people among 4 locations

| Location | User ID | Time |
|----------|---------|------|
| C | 1 | $t_1$ |
| C | 2 | $t_1$ |
| C | 5 | $t_1$ |
| C | 9 | $t_1$ |
| B | 1 | $t_2$ |
| B | 2 | $t_2$ |
| B | 3 | $t_2$ |
| B | 4 | $t_2$ |
| B | 6 | $t_2$ |
| B | 7 | $t_2$ |
| B | 8 | $t_2$ |
| D | 1 | $t_3$ |
| D | 2 | $t_3$ |
| D | 5 | $t_3$ |
| A | 1 | $t_4$ |
| A | 2 | $t_4$ |
| A | 6 | $t_4$ |
| A | 7 | $t_4$ |
| A | 10 | $t_4$ |

Figure S1 shows a simple bipartite graph model of the 4 locations and 10 people. It is obvious that the bipartite model can better reflects the interactions between locations and people.

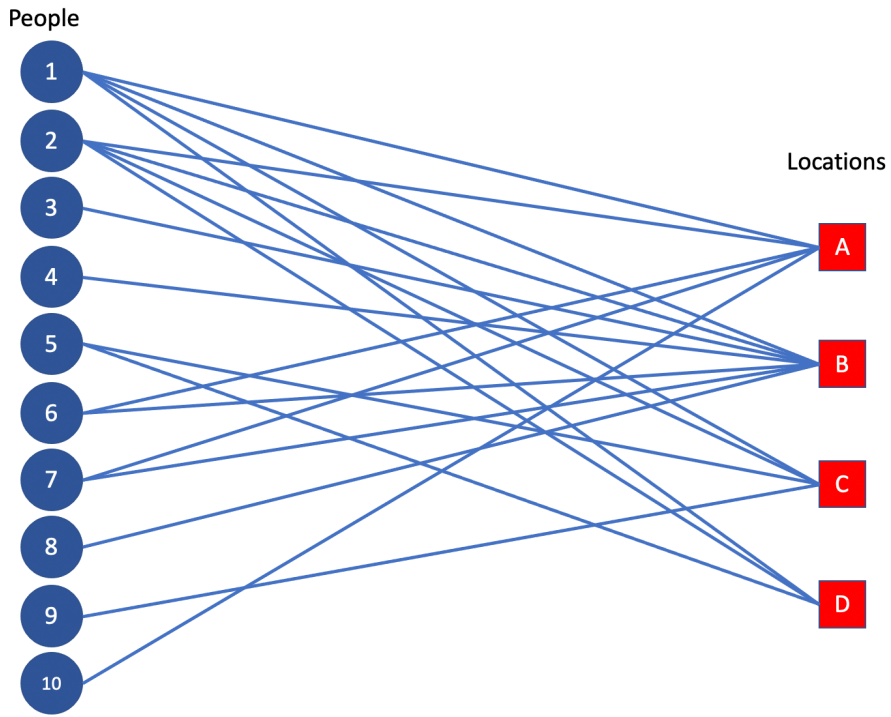

Figure S1 Bipartite graph for 10 people and 4 locations

Table S2 shows the visit history of 5 locations and the 20 visitors of those locations.

Table S2. Locations and visitors

| Locations | Visitors |
|---|---|
| A | 1,2,3,4,5,6,7,10,11,12,14,15,16,17,20 |
| B | 1,2,3,4,6,7,8,11,13,15,19 |
| C | 1,2,5,9,11,12,13,14,16,17,18,19,20 |